# Agentic AI and Machine Learning for Accelerated Materials Discovery and Applications

Jihua Chen[1], Panagiotis Christakopoulos[1], Karuna D. Chen[2], Ilia N. Ivanov[1], Rigoberto Advincula[1,2] *

[1]Center for Nanophase Materials Sciences, Oak Ridge National Laboratory (ORNL)
1 Bethel Valley Road, Oak Ridge, TN 37830

[2]Department of Chemical and Biomolecular Engineering,
University of Tennessee at Knoxville
1512 Middle Dr, Knoxville, TN 37996

This manuscript has been authored by UT-BatteIIe, LLC, under Contract No. DEAC05-00OR22725 with the U.S. Department of Energy. The United States Government and the publisher, by accepting the article for publication, acknowledges that the United States Government retains a nonexclusive, paid-up, irrevocable, world-wide license to publish or reproduce the published form of this manuscript, or allow others to do so, for United States Government purposes. DOE will provide public access to these results of federally sponsored research in accordance with the DOE Public Access Plan
(http://energy.gov/downloads/doe-public-access-plan).

* To whom correspondence should be addressed: radvincu@utk.edu

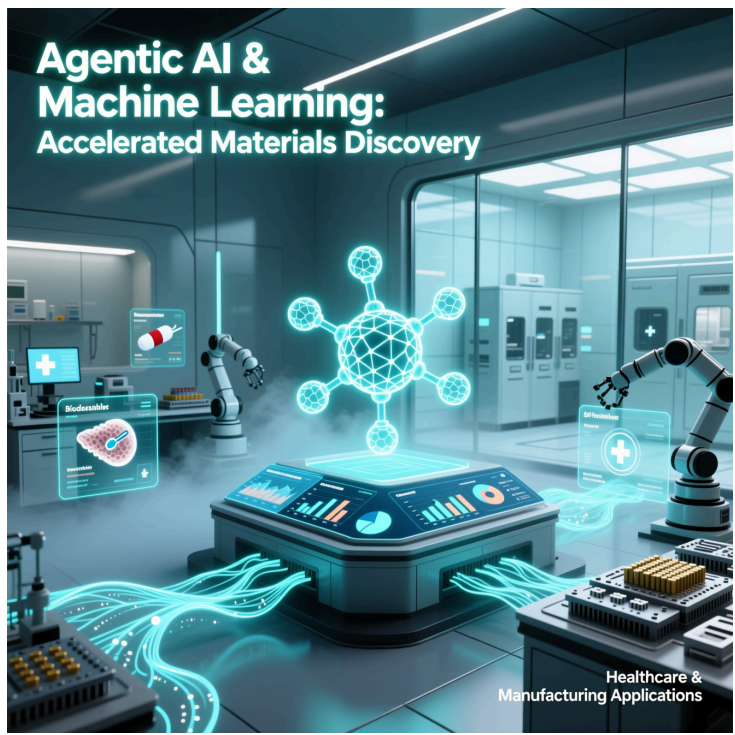

# Abstract

Artificial Intelligence (AI), especially AI agents, is increasingly being applied to chemistry, healthcare, and manufacturing to enhance productivity. In this review, we discuss the progress of AI and agentic AI in areas related to, and beyond polymer materials and discovery chemistry. More specifically, the focus is on the need for efficient discovery, core concepts, and large language models. Consequently, applications are showcased in scenarios such as (1) flow chemistry, (2) biosensors, and (3) batteries.

Table of Content

# 1. Introduction

## 1.1 Need for faster discovery

Historically, materials discovery has relied on experimental validation, characterized by laborious trial-and-error, often resulting in development timelines of 20 years or more.[1] The chemical space, estimated to exceed $10^{60}$ carbon-based molecules, is too vast for exhaustive empirical exploration.[2] Technological advancements (e.g., electric vehicles, aerospace) require new materials with extreme or customized properties, such as ultra-light alloys, high thermal conductivity, or customized band gaps. AI/ML in new materials discovery and chemistry will play a vital role in digital twins, algorithm development, and self-driving laboratories (SDL).[3] For example, in the area of additive manufacturing (AM), AI/ML has a vital role in the discovery of new composites and optimization of the manufacturing method.[4–6] The introduction and demonstration of many SDL capabilities can only accelerate science discovery and hypothesis validation.[7] Computation and simulation from atomistic to coarse-graining and molecular dynamics (MD) methods are important for predictive digital twins; experiments are still needed for validation. While initial computational approaches (e.g., Density Functional Theory, DFT) enable high-throughput identification of candidates, the subsequent experimental realization and validation remain a persistent bottleneck.[8]

## 1.2 AI/ML shift and scope of paper

Machine Learning (ML) has emerged as an efficient surrogate model for accurate property predictions across diverse chemical spaces, circumventing the limitations of *ab initio* calculations. This field, known as Materials Informatics (MI), utilizes computational power to extract unique insights from large datasets.[9] The primary goal is shifting from the traditional experiment-first approach to an AI-driven approach focused on inverse design—determining the structure or composition required to achieve specific desired properties.[10] Agentic AI systems represent the next evolution, operating with autonomy to perform tasks like hypothesis generation, experimental design, and automated data analysis.[1] In this review, we discuss the progress of AI and agentic AI in areas related to and beyond polymer materials and discovery chemistry.(Figure 1) More specifically, the focus is on the need for efficient discovery, core concepts, and large language models. Consequently, applications are showcased in scenarios such as (1) flow chemistry, (2) biosensors, and (3) batteries.

## 2. Core Methods

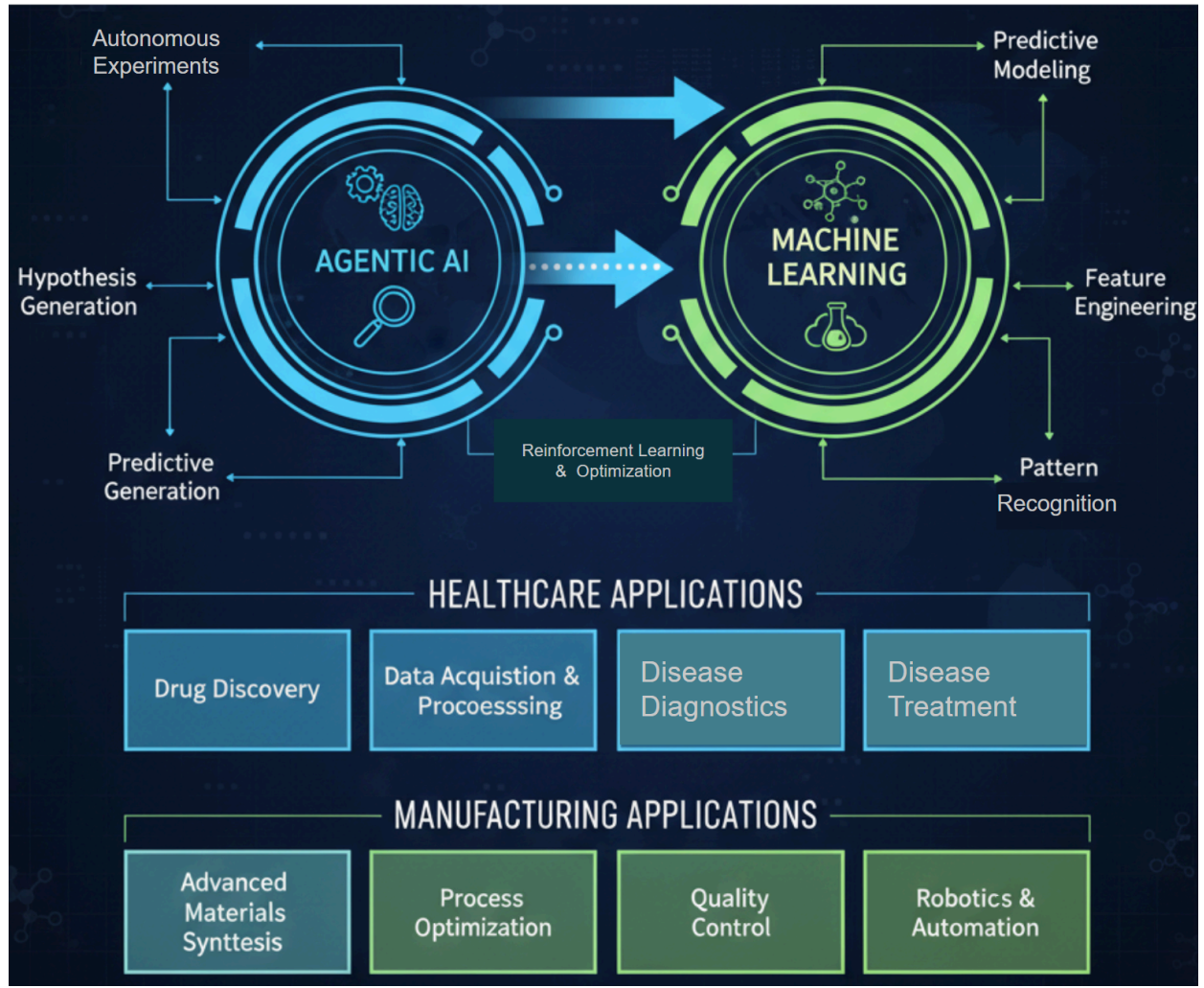

**Figure 1.** AI agents and machine learning for materials discoveries in healthcare and manufacturing

## 2.1 Descriptors and Models

Material Descriptors, as numerical representations, serve as the bridge between AI/ML algorithms and materials chemistry, encoding structure into a machine-readable format.[11] Ideal descriptors must be reproducible, efficient, accurate, and simple. (Figure 2)

The encoding type significantly impacts model success.[10,12] Sequence-Based encoding includes SMILES/SELFIES, graph-based encoding captures connectivity via Graph Neural Networks or GNNs, while multi-modal encoding combines structural, spectroscopic, and text data to improve accuracy.[10,12]

Machine Learning models for chemistry, materials, and manufacturing have been summarized elsewhere.[8] Here we elaborate on Reinforcement Learning (RL) and Deep Learning (DL). Unlike traditional ML, which relies on manual feature engineering, DL uses deep neural networks (DNNs) to autonomously learn hierarchical representations from raw data.[11] RL trains agents in virtual environments, maximizing rewards based on actions and outcomes.[8] This is used to simulate complexity in real-world chemistry experiments, such as predicting optimal synthesis schedules and managing the cost of observation (measurement).

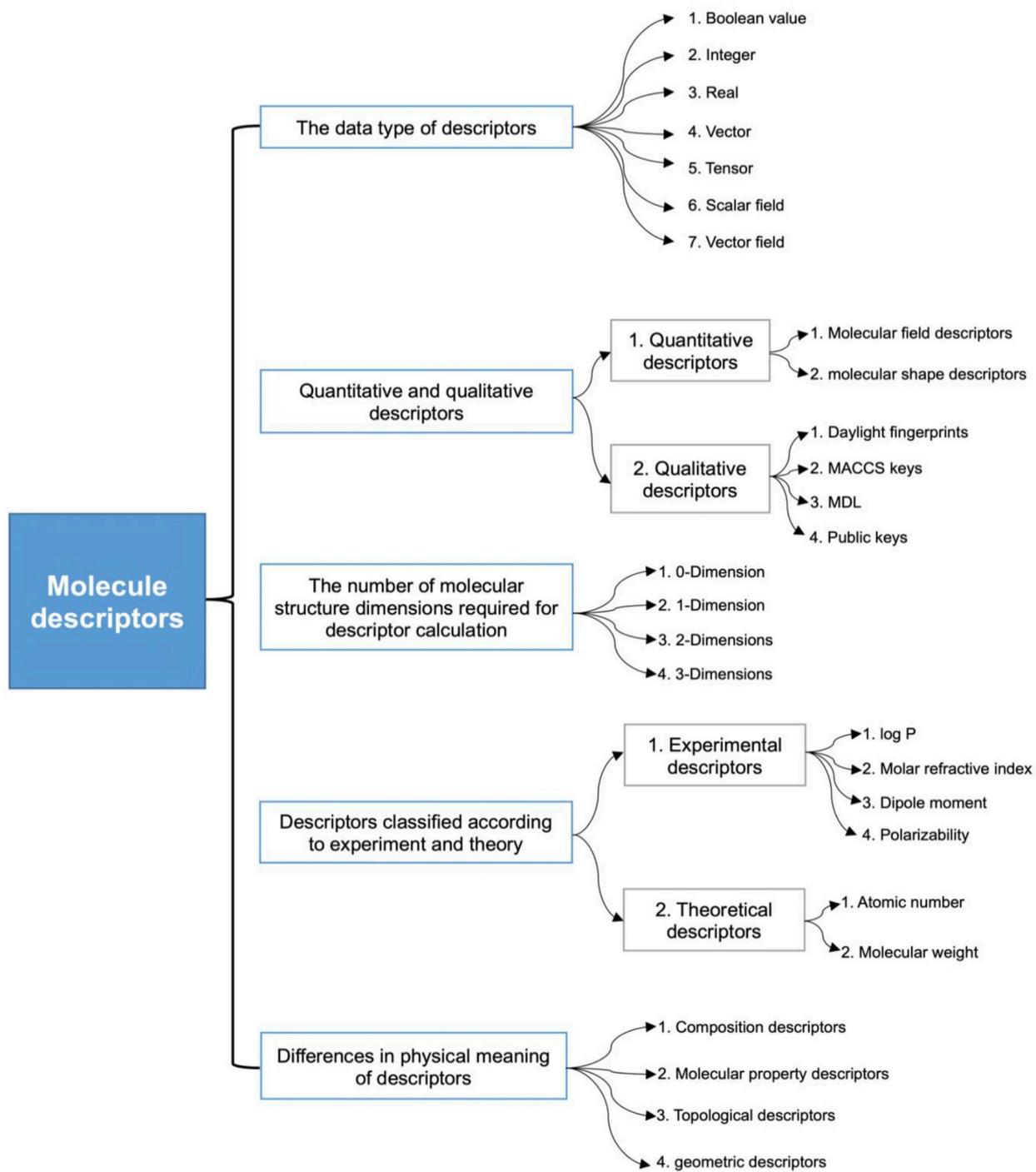

Figure 2. Different types of molecular descriptors. (Figure and Caption from reference,[11] without change, under Creative Commons 4..0 license.[13])

## 2.2 Generative Models and Inverse Design

Generative Models aim to directly generate novel material structures that satisfy specific properties, rather than just screening existing databases.[10,14] Key models include Variational Autoencoders (VAEs), Generative Adversarial Networks (GANs), Diffusion Models, Transformers (e.g., MatterGPT), and Normalizing Flows (NFs).[10] Diffusion Models employ a noise-to-data denoising process, offering high novelty and symmetry preservation, but requiring significant computational resources.[10] MatterGen, for instance, is a Diffusion Model trained on vast stable materials databases (Materials Project) capable of generating materials guided by chemical, mechanical, or electronic property prompts.[14]

## 2.3 Agentic AI Architecture and Autonomy

Agentic AI or AI agent aims to perform a specific task intelligently and continuously on behalf of a human operator.[1] Agentic AI systems rely on an architecture that supports core agency factors: intentionality (planning), forethought, self-reactiveness, and self-reflectiveness. For reliability, autonomy must be balanced with predictability. Orchestration layers handle known procedures (if/then logic, loops, retries), moving them out of the agent's prompts. Agents are generally given a narrow scope of responsibility and perform single, explicit functions for easier testing and debugging.[1]

Explicitly showing the agent's planning steps enhances transparency and debuggability.[15] Frameworks like Curie are specifically designed to embed rigor into scientific experimentation via an ***Experimental Rigor Engine*** that enforces methodical procedure, reliability, and interpretability.[15]

Complex, multi-domain problems benefit from multi-agent setups where specialized agents can collaborate with each other (e.g., planner, executor, evaluator).[1] Examples of AI agents for healthcare and manufacturing include Coscientist[16] for autonomous chemical experiments and CellAgent[17] for bioinformatics data analysis. Table 1 highlights some of the current agentic ai techniques adopted for material discovery and applications, including the use of large language models (LLMs).

Table 1. Current agentic ai techniques for material discovery and applications.[18–22]

| Name | AI Agent Design | Applications |
|---|---|---|
| MOFGen | Collaborative agents for metal organic frameworks (MOF) discovery: a LLM to propose compositions, a diffusion model to generate structures, a quantum mechanical agent for screening, and a final agent to assess synthesizability. | Accelerating discovery for $CO_2$ capture and water harvesting. |
| Eunomia | The LLM-based agent uses the ReAct architecture, enhanced with document search, chain-of-verification, and dataset search, to extract materials entities, relations, properties from literature. | Creating datasets for predictive modeling in materials design |
| LLMatDesign | An LLM agent that performs iterative material modification through prompting, reasoning, and tool invocation to optimize properties. | Enabling discovery for desired electronic, mechanical, etc. in alloys and polymers. |
| MatPilot | An agentic AI for designing experimental schemes, integrating with robotic systems for verification in closed-loop autonomous labs. | Autonomous synthesis and optimization of materials |
| ChemCrow | An LLM-based agent for planning synthesis pathways and designing materials. | Chemical synthesis, and material discovery |
| Aitomia | A computational chemistry agent using ML models via natural language queries, for energy calculations, geometry optimizations, and molecular dynamics. | Predicting properties for energy storage, catalysts, and reaction thermodynamics. |
| Dreams | A multi-agent framework for density functional theory (DFT) simulations, including planning, structure generation, convergence testing, and high performance computing scheduling. | High-throughput materials discovery, such as battery materials like lithium cobalt oxide. |
| A-Lab | An autonomous laboratory system employing AI agents for robotic synthesis, active learning, and decision-making in material optimization. | Robotic synthesis with high success rates |

# 3. LLMs for knowledge mining

The recent addition of reasoning capabilities made these multimodal LLMs a desirable engine for specialized scientific agents, with the functionality of data extraction, multidimensional reaction modeling, experimental workflow optimization, results fitting, and testing enhancement for novel materials.[23,24] The role of these agents is to enable accelerated R&D efforts through the automation of time-consuming manual tasks:

(1) LLMs can extract synthesis conditions from many publications, accomplishing challenging tasks of sieving through the text and data in minutes, also formatting extracted information for future analysis. LLMs can combine information extracted from multiple sources (publications, patents, websites) to build a comprehensive, structured knowledge database covering reactants, reaction conditions, experimental design, product separation, and their properties. Efficient assembly of extensive data from different sources could enable the identification of the range of reaction conditions and establishment of correlations between a product structure and functional properties.

(2) LLMs can facilitate translating complex datasets into actionable reaction protocols (computer commands) for synthesis. This integration can also enable predictive modeling, in which LLMs use prior experimental outcomes to propose novel material compositions and environmental conditions for future testing, ultimately accelerating material discovery and characterization workflows.

Open information extraction toolkit for chemistry literature (such as OpenChemIE) does extraction in two steps: (1) it parses molecules or reactions from the text or figures using specialized neural models, and (2) uses chemistry-informed algorithms to integrate such information and allow for the extraction of fine-grained reaction data from reaction condition and scope investigations.[25] The OpenChemIE performance demonstrated an F1 score of 69.5% with an accuracy score of 64.3% when compared to the Reaxys database.

Those agents promise to streamline synthesis-characterization protocols in the near future. Flow synthesis offers significant advantages over traditional batch synthesis by providing enhanced control, efficiency, and safety (with smaller quantities of reactants).[26] Continuous flow reactors enable precise regulation of reaction conditions, such as temperature, pressure, and flow rate, efficient reactant mixing and heat transfer, ensuring high yield and rate of product generation along with high reproducibility.[27] *In situ* diagnostics of in-line characterization techniques (optical spectroscopy, NMR, IR, HPLC, viscoelastic) allow for immediate adjustments, minimizing side products and enhancing product purity.

Integrating flow synthesis with large language models (LLMs) can assist in designing synthetic routes by analyzing large datasets of reaction conditions, predicting optimal parameters, and, most interestingly, suggesting alternative pathways that have never been tested experimentally.[28] The selective synthesis conditions could be translated into executable autonomous synthesis protocols, streamlining experimental planning and execution. Integration of LLMs within AI/ML centric feedback diagnostics can serve several functions, from identification of reaction state, adjustment of reaction parameters to optimize yield, rate and selectivity, along with making decisions easily interpretable using natural language processing (NLP).

# 4. Applications

## 4.1 Flow chemistry

AI and ML have been proved to be powerful tools for reaction prediction and new materials design. One area that has greatly benefited is the pharmaceutical industry, where AI/ML can be used for the discovery and design of new drugs.[29] Machine learning, and specifically deep learning (DL), has also been broadly used in Analytical chemistry. DL algorithms have been successfully applied to many tasks such as data interpretation, classification, calibration, preprocessing, peak evaluation, and peak picking.[30,31]

*Radical polymerization*

In free radical polymerizations, the reaction is usually performed in bulk. However, this would cause many issues in the small-diameter tubing of the flow reactors due to the high viscosity of the products. Nevertheless, quite a few polymerizations have been conducted in solutions under flow, with one of the first attempts reported in 2005 by Iwasaki *et al.*, who used AIBN to synthesize polymers of butyl acrylate, methyl methacrylate, styrene, benzyl methacrylate and vinyl benzoate.[32] Other monomers have also been polymerized in flow reactors like acrylamide[33] and methacrylic acid,[34] while also studying the kinetics of the reaction. SDL Capabilities will be important for investigating hypothesis driven approaches in copolymerization at a faster speed, ideally with digital twins beyond the Mayo-Lewis Equation.[35]

*Cationic polymerization*

Typically, a highly exothermic polymerization method, cationic polymerization has benefited from the use of flow reactors for the polymer synthesis. Usually, the combination of low temperatures and high flow rates (short residence time) can improve the living character of the polymerization, significantly reducing the polydispersity of the polymers. Isobutyl vinyl ether, n-butyl vinyl ether, tert-butyl vinyl ether,[36] isobutylene[37], and diisopropylbenzene[38] are among the monomers that have been polymerized in flow reactors.

*Nitroxide-mediated polymerization*

Nitroxide-mediated polymerization (NMP) is another example of a polymerization that is conducted at very high temperatures (> 100ºC). The improved heat transfer in the flow reactor over the batch reactor enables more effective control of these polymerizations. Reports from NMP polymerizations show polymers with a narrower distribution compared to batch reactions.[39,40]

*Reversible addition-fragmentation chain transfer*

One of the most versatile polymerization methods is the Reversible addition-fragmentation chain transfer (RAFT) polymerization. The addition of the Chain Transfer Agent into a typical free radical polymerization offers control over the reaction, providing polymers with a narrow molecular weight distribution. The first RAFT polymerization in flow reactors was conducted for the synthesis of poly(N-isopropyl acrylamide) PNIPAM[41], and since then, many monomers and CTAs have been used in polymerizations under flow.[42] Different approaches of RATF have been also discussed, like photoinduced electron/energy transfer reversible addition–fragmentation chain transfer (PET-RAFT)[43] where the initiation step is achieved with light and the absence of radicals produced by chemical initiators.

It becomes clear from the above, that flow reactors can retain the living/ controlled character of polymerization techniques developed over the years and also improve them. Thus, reactions in flow have also been considered to synthesize block copolymers or even polymers with complex macromolecular architectures.[44–47] AI and ML guided flow chemistry have the potential to automate and efficiently optimize such polymerizations. Table 2 highlighted recent works of AI aided flow synthesis of organics and polymers.

Table 2. Current machine learning and ai techniques for flow synthesis of organics and polymers [48–55]

| Algorithm | Approach | Applications |
| --- | --- | --- |
| Bayesian Optimization | A probabilistic method that models uncertainty to efficiently optimize parameters by selecting the next best experiment | Multi-objective optimization of flow synthesis conditions for maximizing polymer yield, monomer conversion, and minimizing cost |
| Thompson Sampling Efficient Multi-Objective Optimization (TS-EMO) | An algorithm using Thompson sampling for exploring Pareto fronts in multi-objective trade-off analysis. | Closed-loop automated optimization of polymer properties like molar mass dispersity and functionality in flow chemistry |
| Active Learning | A strategy that queries the most informative data points to improve model performance, often using uncertainty estimates. | High-throughput exploration of polymer design spaces in flow systems, optimizing synthesis parameters and reducing trials. |
| Reinforcement Learning | Agent-based ML where policies are learned through rewards for optimal decision-making in dynamic environments. | Process control and optimization in flow polymerization for consistent polymer quality. |
| Computer-Aided Synthesis Planning (CASP) with ML | AI-driven planning of synthesis routes using ML for reaction outcome prediction and pathway optimization. | Robotic flow platforms for scalable polymer synthesis, automating multi-stage processes from ideation to execution. |

## 4.2 Biosensors

The development of next-generation biosensors, often relying on micro- and nanofabrication technologies, has led to a broad spectrum of advanced applications in medicine.[56] However, to achieve widespread clinical translation, further research efforts with AI incorporation are expected to overcome a number of critical technical challenges, such as specificity, sensitivity, selectivity, stability, variability, capacity, cost, system integration, as well as sample collection.[56–59]

Next-generation biosensors are being primarily developed for real-time and personalized monitoring of human health, which includes, but is not limited to the following:[56–59]

1. **Remote Monitoring and Point-of-Care (POC)**: Biosensors, particularly wearable sensors (such as skin-integrated systems) or implantable devices, are crucial for monitoring patients with chronic conditions in home and community settings.[56,57,59] For example, highly commercialized glucose sensors are essential for monitoring and controlling blood sugar levels.[59] Researchers aim to achieve non-invasive glucose monitoring via optical or transdermal approaches.[56] Cardiovascular disease sensors are used to monitor and control cardiovascular conditions in real time during and after surgical procedures, aiding in the early detection, monitoring, and treatment of cardiac problems.[60,61] Vital signs and biophysical sensors monitor critical parameters over long periods, including blood pressure, blood glucose level, heart electrical signals, pulse rate, and respiration rate.[62] Miniaturized biosensors can be integrated into personalized point-of-care devices for reliable, sensitive, and fast detection of biomolecules, leveraging their portable features.[63] Currently, there is much attention on the development of wireless electrochemical biosensors, which would increase practicality and convenience for the patient.[62]

2. **Rapid Diagnostics and Therapeutics**: Biosensors are ideal for rapid diagnosis and timely initiation of targeted treatment for infectious diseases (such as HIV, Malaria, Tuberculosis, Sepsis, and Influenza).[64] They can offer a platform for rapid antimicrobial susceptibility testing, which is crucial for combating multi-drug resistant pathogens. Early cancer diagnostic sensors are used to detect cancerous cells and monitor biomarkers associated with the disease.[65] Optical biosensors are useful for diagnosing various diseases in real time due to their fast application and high sensitivity.[64] Electrochemical biosensing systems are being developed to detect nucleic acid and protein biomarkers,[65] which potentially offers an accessible and inexpensive option that efficiently relays diagnostic information. These sensors can be specialized to detect potentially threatening mutations and damages to DNA, and be designed for early detection of cancers.[65] Highly sensitive biosensors with the ability to detect extremely low levels of antigen are particularly suited to detect infectious diseases in their earliest stages.[64]

3. **Biomimetic Sensing**: Biomimetic sensors utilize synthetic receptors and various nanostructures to replicate characteristics similar to those of the human or mammalian systems (such as olfactory and taste receptors).[66] Some biomimetic sensors are designed with highly sensitive, molecular imprinted architecture to best enhance molecule recognition and absorption in order to reflect the complexity of the system in the actual organism.[66] For example, nanofabricated electronic nose sensors (Enoses) including those made with semiconductor devices can be used to detect and analyze various gas contents and volatile organic compounds (VOC).[66,67] (Figure 3 Top) Nanofabricated olfactory and gustatory biosensors can be used to perform biorecognition effectively, when combined with signal processing and pattern recognition techniques. (Figure 3 bottom).[66] Meanwhile, the development of biomimetic membranes create a stable environment optimal for protein function.[66] Synthetic bioreceptors on the surfaces of these membranes present opportunities for biosensor integration, with potential applications in clinical diagnosis and enhanced drug delivery.[66] In the emergence of new infectious diseases, these biomimetic biosensors have the potential to tackle the complexity of unknown and contribute effectively to the monitoring of disease transmission.[66]

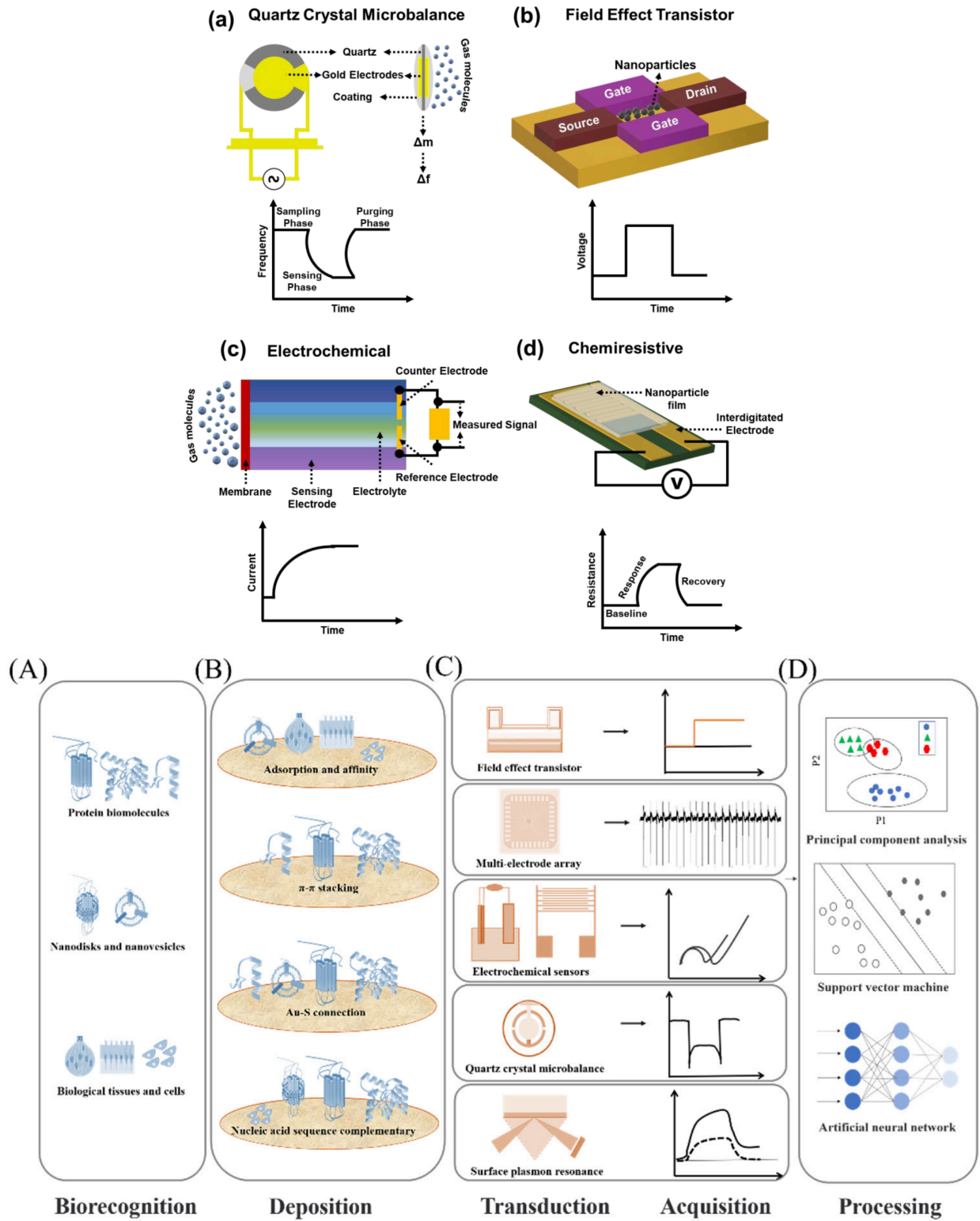

**Figure 3. Top:** Simplified schematic illustrations of different transduction mechanisms utilized in Enoses systems: (a) QCM, (b) FET, (c) electrochemical, (d) chemiresistive. (Figure and Caption from reference[67] without change, under CC-BY 4.0 License.[13]) Bottom: Scheme of main components construction of different detection strategies development for olfactory and gustatory recognition. (A) Biorecognition

parts for the sensing of olfactory and gustatory ligands. (B) Effective deposition methods of sensitive materials on transducers surface. (C) Various transduction techniques to sense the responsive signals. (D) Signal processing and pattern recognition methods. (Figure and Caption from reference,[66] without change, under CC-BY 4.0 License.[13])

Artificial intelligence (AI) and machine learning (ML) are being leveraged throughout the biosensor development life cycle, including biomarker discovery, the physical fabrication, and design optimization of the sensing platform (Table 3).[57,68–71] The following highlights some of the current important cases of AI in biosensors.

1. **Discover biomarkers:** A highly sensitive system coupled with ML algorithms, specifically Random Forest (RF) and Neural Network (NN), demonstrated high precision in prostate cancer screening using just a single drop of urine.[57] AI/ML applications are utilized for biomarker discovery in Alzheimer's disease (AD) and dementia, which currently lack robust biomolecular indicators.[68] For instance, a biosensor based on a programmable curved plasmonic nanoarchitecture was proposed for the precise profiling of exosomal biomarkers for the clinical diagnosis of AD.[69] A point-of-care serodiagnostic test for early-stage Lyme disease utilized a multiplexed paper-based immunoassay combined with machine learning.[69] The ML model helped streamline the process of selecting relevant antigens on the optically analyzed device.[70] An integrated approach combining surface-enhanced Raman spectroscopy (SERS) with a specialized Deep Learning (DL) algorithm called CoVari was proposed for the rapid detection of SARS-CoV-2 variants, capable of simultaneously predicting both viral species and concentrations.[57,71] Additionally, an Artificial Intelligence Nanopore system, which combines ML with nanopore construction, has been developed for Coronavirus detection.[68,71]

2. **Optimize material , transduction, and sensitivity**: For instance, AI is used to *de novo* design metamaterials (materials with specific surface structures) by tailoring their optical properties for biosensing.[68,71] A notable example involves a Deep Learning model capable of predicting plasmonic nanostructure geometries that yield optimal responses upon interaction with various biomarkers.[68] ML approaches were used to train a pressure biosensor developed using the interfacial integration of 2D MXene and 1D graphene nanoribbons.[57,71] The sensor successfully utilized ML to detect various sitting postures with over 95% accuracy. In addition, biomimetic electronic noses are essential for monitoring complex gas mixtures, but the accuracy of these electronic noses is limited with regression algorithms. For better air pollution or VOC detection, a fusion network model was proposed, using a transformer-based multikernel with a CNN/LSTM network, to accurately predict concentrations of $CO$, $SO_2$, $NO_2$ in mixtures, especially in low concentration regimes.[72]

3. **Optimize microfluidics and Lab-on-Chips**: Algorithms such as feedforward neural networks and multi-layer recurrent neural networks (RNNs) are used to analyze data from microfluidic impedance flow cytometry, which classify red blood cells from ghost cells with high accuracy (e.g., 96.6%) by interpreting electrical impedance measurements.[70] Reinforcement learning is applied to automate the operation of fluidic control components, thus enabling precise control of flow rates and particle manipulation.[73] ML assists in developing multi-sensor platforms for complex systems, such as the vascularized Kidney-on-Chip, which utilizes biosensors to evaluate nephrotoxicity.[73]

Table 3. Current machine learning and AI techniques for biosensors[74–77]

| AI Technique | Approach | Applications |
| --- | --- | --- |
| Support Vector Machines (SVM) | High-dimensional classification & regression | Pathogen detection, disease diagnosis, biomarker classification |
| Random Forest / XGBoost | Ensemble methods for accurate classification & feature importance | Bacteria discrimination, multiplexed detection, EIS signal analysis |
| Convolutional Neural Networks (CNN) | Automatic feature extraction from spectra/images | Rapid pathogen ID (SERS, fluorescence), optical biosensor analysis |
| Artificial Neural Networks (ANN) | Non-linear pattern recognition | Glucose/lactate sensing, pathogen classification in arrays |
| Principal Component Analysis (PCA) | Dimensionality reduction & pattern discovery | Noise reduction, clustering in sensor arrays & colorimetric detection |
| Reinforcement Learning | Real-time optimization through trial-and-error | Dynamic calibration, adaptive control in wearable biosensors |
| Natural Language Processing (NLP) | Knowledge extraction from scientific literature | Biomarker discovery, automated database creation for sensor design |

## 4.3 Batteries

A core application of AI is the dramatic acceleration of the typically slow and resource-intensive process of finding new battery chemistries (Figure 4).[78] For example, a machine learning model named DRXNet[79] was developed to encode and learn (electro)chemical information from voltage profiles. This model was trained on a dataset of 19,000+ discharge voltage profiles collected in more than five years, from 14 metal species within the Disordered Rocksalt (DRX) cathode chemistry space. DRXNet captures important features in cycling curves across various conditions, offering a data-driven approach to facilitate rapid cathode materials discovery.

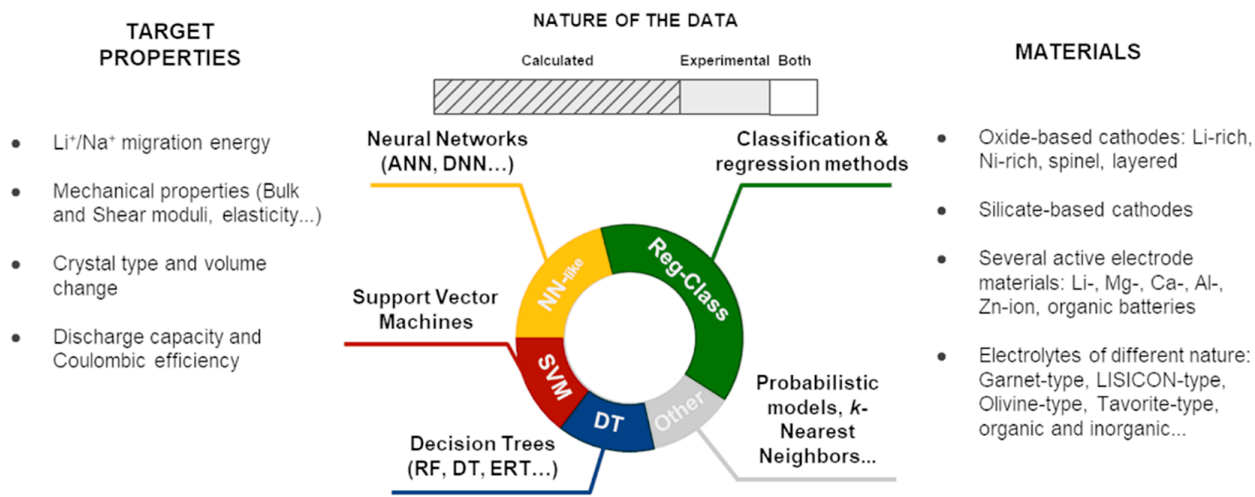

**Figure 4.** Infographic on the ML methods recently used in the literature to search for new battery materials with specific target properties, including the corresponding nature (calculated vs experimental data) of the employed databases. (Figure and Caption from reference,[78] without change, under CC-BY 4.0 NC ND License.[80])

AI is also used to predict battery life accurately using only early-cycle data, dramatically shortening the required testing time.[81–83] Comprehensive experimental data were combined with algorithms such as Bayesian optimization to achieve such goals.[81,83] A machine learning model was trained with a few hundred million data points of batteries charging and discharging.[81] The algorithm could then predict how many more cycles each battery would last based on voltage declines and a few other factors observed during the early cycles. In addition, a closed-loop optimization system compressed the original experimental period of over 500 days to just 16 days when applied to optimizing fast-charging protocols.[83] Furthermore, an ensemble learning framework based on the RUBoost ML method was proposed for classifying electrode quality during manufacturing, and accurately classified three critical quality indicators (electronic conductivity, thickness, and half-cell capacity) for $LiFePO_4$- and $Li_4Ti_5O_{12}$- based electrodes.[84] AI/ML also provides new ways to address the challenges of mass-scale recycling and automated disassembly of Electric Vehicle Lithium-Ion Batteries (EV-LIBs) via a system using anomaly detection, predictive estimation, and high throughput testing.[85]

Table 4 summarizing current machine learning and AI techniques for battery developments, in categories spanning from active learning and reinforcement learning to deep learning, natural language processing (NLP), and Generative AI.

Table 4. Current machine learning and AI techniques for battery developments.[78,86–93]

| AI Technique | Description | Applications |
| --- | --- | --- |
| Supervised ML (SVM, RF, XGBoost) | Algorithms for prediction and classification from labeled data | State Of Health(SOH) /Remaining Useful Life (RUL) estimation, degradation prediction, material optimization |
| Deep Learning (ANN, CNN, LSTM) | Multi-layer neural networks for feature extraction and time-series analysis | Cycle life prediction, microstructure analysis, dendrite/SEI simulation |
| Reinforcement Learning | Trial-and-error learning for optimal decision-making | Charging protocol optimization, battery management systems |
| Generative Models (GAN, VAE) | Creation of novel structures and chemical compositions | New electrode/electrolyte design, high-throughput material screening |
| Hybrid AI-Physics Models | ML combined with physical/electrochemical equations | Long-term performance prediction, interface dynamics simulation |
| Active Learning & Bayesian Opt. | Iterative, uncertainty-guided experiment selection | Efficient material synthesis optimization, reduced experimental trials |
| Natural Language Processing | Text mining and knowledge extraction from literature | Database building, trend identification, research acceleration |

## 5. Outlook

In summary, Agentic AI and machine learning will be important for a wide range of materials discovery and applications in healthcare and manufacturing. Here we used flow chemistry of polymers, biosensors, and battery developments as examples. We expect there will be a number of challenges and opportunities towards the development of AI for future materials discovery and applications in healthcare and manufacturing.

(1) The future direction involves the full integration of computation, ML, and robotics into fully autonomous laboratories that operate 24/7 with minimal human input. The democratization of SDL and ML-driven manufacturing methods such as 3D printing will be a game changer.[94,95] This can largely automate and accelerate materials discovery, in sensor development, composite materials, and beyond.[96–99]

(2) A critical challenge is ensuring the trustworthiness and reliability of AI agents. Autonomous agents are prone to hallucination or operational errors, and need sophisticated oversight. As agents gain autonomy and integrate with physical robotic systems, the potential for compounding errors, the generation of hazardous materials, and alignment failure increases. Robust AI governance, ethical guidelines, and maintaining human oversight are paramount to mitigating these risks.

(3) Future research will continue to tackle generalization across diverse chemical and material spaces. Strategies include developing standardized benchmarks, expanding datasets (via federated learning or synthetic data generation using Diffusion Models), and implementing physics-informed models to ensure realistic outputs.

# Funding and Acknowledgement

This work was supported by the Center for Nanophase Materials Sciences (CNMS), which is a US Department of Energy, Office of Science User Facility at Oak Ridge National Laboratory, and Laboratory Directed R&D (ORNL INTERSECT). KDC is a research intern at the University of Tennessee at Knoxville and a student at Farragut High School.

# Conflicts of Interest

The author declares no conflict of interest.

# Author Contributions

RCA planned the layout and drafted the introduction and Section 1. JC contributed to the overall design and writing. KDC contributed to the Biosensor section. PC, INI drafted Section 4.1 and 3.

Late Update: 1/13/2026